\documentclass[sigconf, table]{acmart}
\usepackage{multirow}
\usepackage{fancyhdr}
\AtBeginDocument{%
  }

\setcopyright{acmlicensed}
\copyrightyear{2024}
\acmYear{2024}
\acmDOI{XXXXXXX.XXXXXXX}

\acmConference[ACM]{}{2024}{USA}
\acmISBN{978-1-4503-XXXX-X/18/06}

\begin{document}

\pagestyle{fancy}
\fancyhf{}
\fancyfoot[C]{\footnotesize©{Saquib Ahmed, Md Nazmus Sakib, Sanorita Dey| ACM} {2024}. This is the author's version of the work. It is posted here for your personal use. Not for redistribution.}

\title{Exploring the Influence of Online Videos on Parents or Caregivers of Children with Developmental Delays}

\author{Saquib Ahmed}
\email{saquiba1@umbc.edu}
\affiliation{%
  \institution{University of Maryland, Baltimore County}
  \city{ }
  \state{}
  \country{}
}

\author{Md Nazmus Sakib}
\email{msakib1@umbc.edu}
\affiliation{%
  \institution{University of Maryland, Baltimore County}
  \city{ }
  \state{}
  \country{}
}

\author{Sanorita Dey}
\email{sanorita@umbc.edu}
\affiliation{%
  \institution{University of Maryland, Baltimore County}
  \city{ }
  \state{}
  \country{}}

\renewcommand{\shortauthors}{Ahmed et al.}

\begin{abstract}
  Developmental Delays and Disabilities (DDDs) refer to conditions where children are slower or unable to reach developmental milestones compared to typically developing children. This can cause significant stress for parents, leading to social isolation and loneliness. Online videos, particularly those on YouTube, aim to support these parents and caregivers by offering guidance and assistance. Studies show that parents of children with DDDs create videos on YouTube to enhance authenticity and build connections. However, there is limited knowledge about how other parents with children with DDDs perceive and are impacted by these videos. Our study used a mixed-method approach to annotate and analyze more than fifteen hundred YouTube videos on children's DDDs. We found that these videos provide crucial informational content and offer mental and emotional support through shared personal experiences. Comments analysis revealed a strong sense of community among YouTubers and viewers. Interviews with parents of children with DDDs showed that they find these videos relatable and essential for managing their children's diagnosis and treatments. We concluded by discussing platform-centric design implications for supporting parents and other caregivers of children with DDDs.
\end{abstract}




\maketitle

\section{Introduction}
Children at an early age go through several types of developmental growth, such as physical, cognitive, language/ communication, social, emotional, or behavioral skills. In each area, children need to reach some milestones at certain ages to be considered typically developing children. However, when a child becomes slow to achieve one or more of these milestones or cannot even reach them compared to other typically developing children, this condition is called developmental delay or disability. Children with Developmental Delays and Disabilities (DDDs) must start early interventions and treatments to make progress or even catch up with their peers~\cite{bax1987medical, tebruegge2004does, oberklaid2005developmental, council2006identifying}. Untreated DDDs can contribute to early school failures and social and emotional problems~\cite{council2006identifying}.

Identifying the condition of children with DDDs is crucial to start early intervention programs for their recovery. However, identifying DDDs, especially among young children, can be challenging for parents~\cite{aap, rosenberg2013computing}. It becomes more complicated when parents, with a busy schedule and multiple stressors, have little to no knowledge of developmental delay and disabilities~\cite{adams2013early, marshall2016parent, woolfenden2015equitable}. Pediatricians and primary care providers should take the lead role in making parents aware of the complications of DDDs. However, early identification and intervention are not widely adopted among pediatricians, even in the USA, despite ongoing education and promotion efforts over many years. Children often do not receive a periodic developmental assessment~\cite{halfon2004assessing, sand2005pediatricians}. Fewer than half of pediatric practitioners use formal screening tools~\cite{radecki2011trends}, and only one-fifth of children received parent-centered developmental screening in a 12-month period~\cite{bethell2011rates}. 

Previous work has shown that parents of children with DDDs often feel more strained and prolonged levels of stress than parents of children without these conditions~\cite{hauser2001children, baker2003pre, spratt2007assessing, lopez2008parental}. Many parents withdraw from friends and family members because they feel embarrassed, anxious, exhausted, and stressed. Parental stress found to have association with lower marital satisfaction~\cite{robinson2015marital}, lower self-esteem~\cite{crnic1984maternal}, lower self-efficacy~\cite{jackson2000parenting}, lower parental well-being~\cite{pisula2007comparative}, and lower satisfaction~\cite{esdaile1995issues, koeske1990buffering} in the parental role. To assist parents in understanding the concept of DDDs and to guide them through the necessary set of actions, experts and domain specialists often choose social media platforms such as YouTube to create video content on children's DDDs to meet informational needs. Parents with experiences raising children with DDDs also share emotionally supporting videos on YouTube with other fellow parents where they talk about their journey, challenges, status, and actions.

In HCI, researchers have explored the affordances of Twitter \cite{elkaim2022deep}, Facebook \cite{gage2017cancer}, and Reddit communities\cite{lee2021using} in supporting patients and their caregivers. Researchers observed how social media helped caregivers overcome loneliness and assisted them in being a part of the community that cares for them~\cite{n2020investigating, johnson2022s}. However, there needs to be more understanding of how YouTube, a social media platform and a relatively passive online community focusing on video content, impacts caregivers. YouTube, the second most popular social media platform~ \cite{ng2023exploring, osman2022youtube}, encourages people, even those with limited experience in video creation, to spread their ideas and knowledge through videos and allows them to share it with a broad audience. On the other hand, the audience often prefers to consume YouTube videos because this platform does not require a personal profile or direct connections~\cite{burgess2018youtube}. Moreover, videos convey a comprehensive set of non-verbal cues such as subtlety, nuance, emotion, sincerity, and gratitude to their audience, which help people passively bond with the content and audience of the content.
To understand the relevance and importance of YouTube videos from the perspective of parents and caregivers of developmentally delayed and disabled children, we asked the following two research questions: 

\begin{itemize}
\item \textbf{RQ1:} What content do people share on YouTube on children's DDDs and how do they present their content to their target audience?
\item \textbf{RQ2:} How do the parents/viewers interact and get impacted by these videos? What do parents intend to learn from these videos? How do these videos impact their actions and (if at all) emotionally support them? 
\end{itemize}

To answer these research questions, we conducted a mixed-method study. To answer RQ1, we annotated and analyzed 1532 YouTube videos on children's DDDs. We identified seven types of video content: videos defining concepts, explaining signs of DDDs, presenting treatments, showing resources and facilities, presenting actionable items, sharing personal experiences, and reporting the recovery journey. In addition, we observed four primary video styles having 11 subcategories. This analysis showed us how developmental experts utilize YouTube as a widely accessible medium to teach parents and caregivers about the symptoms, treatments, and at-home care critical for various forms of DDDs in a way that is easily consumable by non-expert audiences. The findings are more detailed in section ~\ref{req1}.

The second research question aimed to observe how these videos impact the target audience. To this end, we decided to take a two-way approach: 1) analyzing the comments section of the YouTube videos and 2) interviewing the parents and caregivers of children with DDDs who consume these videos regularly. An extensive qualitative analysis helped us identify primarily six types of comments such as 1) asking for suggestions, 2) sharing experiences and concerns, 3) thanking content creators, 4) responses from video creators, 5) sharing comments among fellow video creators, and 6) miscellaneous. Overall, the comments showed how parents and caregivers of children with DDDs receive perceptual, emotional, and informational benefits not only by watching these YouTube videos but also by maintaining an empathizing online community. 

We extended our investigation by interviewing 16 parents who consumed videos related to children's DDDs from Youtube. These interviews revealed how parents consider YouTube videos prepared by DDD experts as one of the primary sources of information at various stages of diagnosis, treatment, and follow-up care of their children. We also learned about the emotional support these video creators provide parents, whereas parents often find it hard to share their challenges with immediate friends and family members. In addition, YouTube allowed parents to remain connected to a significant source of information. At the same time, parents also felt a sense of seclusion, which is not usually experienced in other social media platforms. The findings of the comment analysis and interviews' conclusions are discussed in detail in section~\ref{sci} and ~\ref{fi}. 

It is essential to address that the term "Developmental Delays and Disabilities (DDDs)" is not a medical term. We coined this term to explain our objectives and findings better. We understand that the long-term impact of developmental delays and developmental disabilities can differ significantly. For instance, children with proper and timely interventions can overcome developmental delays completely, whereas developmental disabilities are typically complex to cure completely. In the scope of this paper, we considered those two terms together as an umbrella term because, for young children, the early signs and symptoms of delays and disabilities overlap in many scenarios. Thus, they receive similar interventions and therapies at the early stage. Future work may consider them separately if they focus on inevitable delay or disability conditions.

\section{Related Work}
The related work section explains the challenges and interventions related to children's DDDs. We followed our discussion on how online resources and more importantly, social media platforms impact people's social and emotional state during critical health conditions. 

\subsection{Developmental Delays and Disabilities and Early Intervention Programs}
Developmental delays and disabilities among young children are not uncommon. An "Early Childhood Longitudinal Study" in the USA showed that 13\% of newborn children eligible to receive counseling from early intervention programs developed DDDs at the age of 9 to 24 months. More alarmingly, a comparative study showed that between 1997 and 2008, DDDs among young children increased by 17.1\%~\cite{scherzer2012global}. Yet, only 10\% of those children received services, and this percentage is even lower for black children~\cite{rosenberg2008prevalence}.

The severity of this problem is more intense in low- and middle-income countries (LMICs) as they often lack facilities critical for early identification and intervention of DDDs. Health staff often have limited sensitization, interest, or training in child development or recognition of early delays~\cite{lian2003general}. Apart from health professionals, parents in LIMCs are often unaware of the significance of serious DDDs~\cite{winkvist1997images, ertem2007mothers, williams2007learning}. Usually, medical attention is sought because of acute illness rather than developmental or behavioral concerns. Multiple factors influence the acceptance and practice of early detection and intervention. These include physicians' attitudes, awareness, and interest~\cite{esposito1978physicians}, insufficient training~\cite{thorburn1993recent}, non-acceptance of early treatment~\cite{mousmanis200818}, uncertainty about how and where to refer~\cite{desai2011exploratory}, time limitations of the clinic visit~\cite{ertem2009addressing}, and cost factor~\cite{dobrez2001estimating}. In some cases, practitioners might be legitimately concerned about unnecessarily alerting a family and would prefer to wait until the problem is too obvious to ignore~\cite{shevell2001profile}.    

Despite these challenges, initiatives and trials showed that there are feasible solutions to detect and intervene in DDDs. For instance, WHO's Care for Development and counseling materials use counseling sessions to promote the healthy development of children. Another strategy is to use a surveillance approach in primary pediatric healthcare for routine observation of early childhood milestones~\cite{department2008child, brinkman2014data}. To incorporate parents and children from remote locations in the detection and intervention programs, telehealth is an effective and highly satisfactory method for its stakeholders~\cite{la2022systematic, sengupta2022physician}. Such initiatives work more effectively when parents and caregivers gain familiarity with this topic. One way to promote this familiarity is through YouTube videos on DDDs. In the scope of this paper, we want to explore how YouTube videos on DDDs of various types impact the informational and emotional provisions of the target audience. Here, we discussed, in general, the effect of online resources and social media platforms in the healthcare domain, which will set up the background of this paper.

\subsection{Seeking Health information from Online Resources}
Seeking health information from online resources has gained popularity exponentially. One critical aspect of seeking online health information is verifying the credibility of the data. Studies showed that older adults were often found to be less sensitive to the credibility cues indicated by website features and performed less deliberation to assess the credibility of online health information compared to young users (e.g., checking the source of information)~\cite{liao2014age, berkowsky2018challenges}. Studies found that older adults consider all online health information credible as they believe people are not allowed to post medical information on the internet unless it is checked by someone knowledgeable~\cite{robertson2011exploration}.

Apart from age, the experience of searching for online health information may differ by socio-demographic groups~\cite{prestin2015online, mccloud2016beyond, nguyen2017persistent}. For example, in studies published during the last decade, persons who reported more incredible difficulty seeking online health information were more likely to be from socially disadvantaged groups. These groups are more likely to report negative perceptions about health care than those from relatively advantaged groups~\cite{arora2008frustrated, nguyen2017persistent, miller2012online, finney2019online}. Moreover, because of low literacy, people often got frustrated when they needed help finding their desired health information online, impacting the overall quality of the information they gathered~\cite{mccloud2016beyond}. However, familiarity with technologies does not necessarily imply a less challenging experience gathering online health information. Young, tech-savvy users usually prefer to get information from social media platforms. Health information on social media platforms needs to be more complex, accurate, biased, or a combination. Young users are also prone to take shortcuts when seeking information. Such practice can lead to lower quality information, contributing to increased health anxiety~\cite{colditz2018adolescents}. Despite these challenges, the easy and convenient access to healthcare information from online resources made it a dominant source of information for many healthcare conditions, and this trend became more popular through social media platforms (discussed next).

\subsection{The use of Social Media in Healthcare Domain}
The practice of searching for health information on social media depends not only on the age range (young users) of users but also on the type of information. It was shown that users prefer social media to seek health information on conditions where they expect benign explanations. However, people choose search engines for severe medical conditions over social media~\cite{10.1145/2556288.2557214}. However, the remarkable upsurge of healthcare information on social media changed many existing trends and practices. People should have stayed contained to using social media just for seeking and sharing health information~\cite{fergie2016young, shaw2011health}. They used the medium to communicate with their clinicians and other fellow patients~\cite{naslund2016future, info:doi/10.2196/jmir.1696}. Public health surveillance~\cite{davila2012frequency, jashinsky2014tracking, rosen2013facebook}, disease trend prediction\cite{santos2014analysing, mcgough2017forecasting}, and diseases interventions~\cite{tanner2016wecare, valimaki2016effectiveness} are other dominant areas in healthcare domain where people accessed social media platforms for their benefit.

Despite the wide adoption of social media in healthcare information, scholars have found malicious, suspicious, and misleading information to be spreading on social media. For example, an analysis of tweets during the Zika virus outbreak found a disconnection between the interests and concerns of the general public and the communication of public health authorities~\cite{gui2018multidimensional} and reported widespread uncertainty and ambiguity~\cite{gui2017understanding}. False information created anxiety and panic, which further exacerbated the problem~\cite{raza2022fostering}. Although these findings are alarming, they do not entirely diminish the impact of social media in the healthcare domain. During COVID-19, people use social media to disclose their distress to online connections. Privacy-protecting strategies became the least of people's concerns during stress or crisis~\cite{zhang2021distress}. Studies also found that social media became the primary source of healthcare information during COVID-19, and those who consumed them were more likely to get through vaccination~\cite{neely2021health}. Most existing studies on the healthcare domain utilized text-based social media platforms such as Facebook, Twitter, and Weibo. Little is known about the impact of YouTube videos in this context. Our work examined YouTube videos on children's DDDs with a different lens. We aimed to observe how well the structural components of YouTube videos reflect social provisions and thus create a sense of bonding within the community.

\section{Methodology}

This research aims to understand the characteristics of video content shared on YouTube on children's DDDs and how that content impacts the viewer community. In RQ1, we aimed to analyze these videos' content and presentation styles. To this end, we collected YouTube videos on children's DDDs, and after initial scrutiny, we meticulously annotated them to examine their content and presentation styles. We extended our analysis further by performing qualitative analysis on comments to understand how these videos created a virtual viewers' community and how viewers consume the content of these videos (RQ2). To answer RQ2, we further interviewed parents of children with DDDs. The qualitative analysis of these interview transcripts revealed why parents consume these videos and how the content of these videos impact them in various context. We summarized our findings and the practical implications of our work in the paper's discussion section, reiterating the thoroughness of our analysis and the validity of our research.

\subsection{Data Collection (RQ1 and RQ2)}
We used YouTube Data API~\cite{yt} to crawl the data. We followed an ethical web scraping protocol to avoid automatic bans. Before collecting data, we made developer accounts and were approved for the platform to use the account for collecting data related to academic research. We avoided multiple, fast-paced requests for YouTube API to prevent overloading the platform server. Instead, we set a crawl delay proportional to the time of a regular human speed browsing the platform for videos. We used a list of keywords such as "developmental disabilities in children," "toddler developmental delays/disabilities," "children developmental delays/disabilities," "cognitive delay/ disabilities," "motor delays/ disabilities," "sensory delays/ disabilities," "social delays/ disabilities," "emotional delays/ disabilities," "speech delays /disabilities," and "behavioral delays/ disabilities" along with different combinations of these keywords to crawl the initial list of videos. Video metadata includes the title, description, duration, publishing date, view count, like count, comment count, subscriber count, and comments. Through our initial crawling, we retrieved 6,735 videos. We thoroughly scrutinized all videos in two rounds. In the first round, we eliminated all videos that either did not have any channel information (for authenticity) or were shorter than 10 seconds in length. This elimination process also removed private videos, deleted videos, and videos with broken links. In the second round, we manually coded all videos to eliminate videos unrelated to children's DDDs. For example, videos related to mental health challenges among high school students or videos on general developmental steps among non-delayed children were eliminated in this round. We also eliminated videos that either used more than one language in their presentation or used other than English in their presentation. Two authors started with 100 randomly picked videos to individually determine their list of videos that should be eliminated from the list. Once those individual lists were made, two authors discussed their reasoning for all the videos they decided to delete. When they both agreed with their initial list of rejected videos, they reviewed the remaining videos individually (the first author examined 3,319 videos, and the second author reviewed 3,316 videos from the remaining list). After two rounds of cleaning and elimination, 1532 remained in the dataset we used for analysis. The videos were from 527 unique YouTube channels, viewed 1,976,881,082 times, and received 111020 comments and 11,271,278 likes. The median duration of the videos was 5.65 minutes (mean = 13.94, SD = 17.24 ).

\subsection{Analyzing Video Content and Styles (RQ1)}
\label{lb2}
We performed qualitative annotation of a subset of our dataset to identify the content and production styles of the videos. We used Grounded Theory, iteratively open-coding each video, and then created and refined our annotations. More specifically, all three authors of this paper annotated a random sample of approximately 500 videos (the first author annotated 532 videos) from our dataset. All authors had prior experience in analyzing healthcare-related YouTube videos. None of the authors were doctors. HCI researchers studied children's CDs from resources published by CDC, Mayo Clinic, Kennedy Krieger, and Johns Hopkins Pediatric Developmental Medicine for at least three months before this annotation task. Without labeled ground-truth data, all three authors independently adhered to an open inductive coding approach~\cite{glaser1968discovery}. We organized multiple brainstorming sessions during this coding process, where all three coders discussed their preliminary thoughts, confusions, and opinions. The annotation procedure entailed an initial set of codes to use and then, second, coding each video independently, marking salient themes to represent the categories and styles observed in the videos with our labels. We agreed substantially among three coders based on Fleiss' kappa test (K = 0.89, p $<$ 0.05).

Once all three coders independently finished coding all 1,532 videos, we met as a team to discuss the themes observed and then decide on a refined set of codes. We identified 14 video styles and 12 types of content. Next, we invited five undergraduate students to avoid any bias imposed by the authors of this paper. All of them were in their senior year (three Computer Science majors and two Information Science majors) and had prior experience in annotating social media video content. They examined a random sample of 200 videos (only one video was chosen from each channel to maintain diversity and uniform representation). To provide background on the annotation process, we conducted a two-hour-long information session discussing styles and content identified earlier, along with examples. All coders independently coded this set of 200 videos. They could apply any categories from the existing pool (if applicable) or create a new style or content category for each video based on their judgment. Finally, we discussed their coding experiences and received feedback about potentially ambiguous, misrepresented, and possible new themes. All three authors of the paper met together after the annotation session with undergrad coders and finalized four primary video styles (11 subcategories) and seven types of content through this process.

\subsection{Exploring Viewer Involvement on YouTube(RQ2)}

To understand how viewers interact with these videos, we analyzed the view count, like count, and comments of all videos in our dataset. We could not include the dislike count of the videos as YouTube hid that information from public view starting in November 2021. We calculated the reach and engagement factors for each video using these parameters. YouTube does not disclose the number of unique users who viewed a video. Thus, these platform statistics mentioned above are usually used to examine viewers' engagement and reach~\cite{bartl2018youtube, biel2009wearing, borghol2012untold, niu2021stayhome, chatzopoulou2010first, niu2021teamtrees}. The total view count of videos measures the reach factor of a video. On the other hand, the engagement factor considers both like count and comment count~\cite{hair2017harnessing, loft2020using}. These parameters are used to calculate the like rate (number of likes received per 100 views) and comment rate (number of comments per 100 views), which estimate viewer engagement.

In addition to numerical measures of viewers' reach and engagement, we qualitatively analyzed comments collected from our dataset's videos. Creators turned off the comment property for 121 videos in our dataset. We eliminated those videos from the comment analysis. In the remaining dataset, we only considered comments from videos with at least five comments in their comment section (89 videos). Finally, we processed comments collected from 1322 videos. We performed iterative open-ended coding~\cite{strauss1998basics, strauss1987qualitative}, followed by thematic analysis~\cite{denzin2011sage, miles1994qualitative, patton2002qualitative}. To this end, we followed a similar approach in section~\ref{lb2} to find the main themes from the comments. All three authors of the paper applied the open coding method to identify the first set of themes for 200 randomly chosen comments from our dataset. At most, ten comments were chosen from a single video to maintain variation. Once all three authors identified the first set of themes individually, we discussed the themes we each identified and finalized a list of six themes after two iterations. As before, each author took a new set of 3000 comments and coded them separately to verify the efficacy of those six codes. Once those codes were completed, we measured inter-rater agreement using Fleiss' kappa measure and found almost perfect agreement between all three authors (K = 0.86, p $<$ 0.05). This list covered around 7.74\% comments of the full list of 111028 comments. The analysis helped us perceive how viewers engage with the videos on DDDs and how these videos impact their sense of community bonding and social presence. We briefly discussed all six themes below and an example from each theme (presented in table \ref{tab4}).

\subsection{Interviewing Parents of Children with DDDs (RQ2)}
A critical aspect of observing the significance of YouTube videos on parents and caregivers of children with DDDs is to learn from the parents directly why and how they consume these videos and how those videos have impacted them. To this end, we conducted interviews with sixteen parents whose children have some form of DDDs. Here, it is critical to mention that interviewing parents of children with DDDs is a complex process. The sensitive nature of DDDs requires an overwhelming amount of physical and emotional involvement from the parents' side. Parents often do not feel comfortable discussing this topic with external researchers because of their personal reservations. Understanding the sensitive nature of this issue, we refrained from contacting parents directly. Instead, we reached out to them through private Facebook groups (dedicated to parents of children with various forms of DDDs) by posting announcements with the help of group admins. 22 parents initially contacted us for the interview, but later, 6 of them had to cancel the meeting due to personal issues. All interviews were conducted through video calls. The demographics of the interviewees are presented in table \ref{tab1}.

\begin{table*}[h!]
\centering
\caption{Demographic Information of the Interview Participants}
\label{tab1}
\begin{tabular}{|l|l|l|l|l|l|}
\hline
\textbf{Participant} & \textbf{Education} & \textbf{Race} & \textbf{Age (Child)} & \textbf{DDD(s)}  \\ \hline
P01 & Master's & White & 7 & ADHD \\ \hline
P02 & Bachelor's & Asian & 4 & Speech \\ \hline
P03 & Bachelor's & Black & 17 & Motor \\ \hline
P04 & Master's & White & 8 & ASD \\ \hline
P05 & Bachelor's & Hispanic & 13 & Speech \\ \hline
P06 & Master's & Black & 5 & TBCD \\ \hline
P07 & High School & White & 9 & Autism \\ \hline
P08 & Master's & Asian & 12 & CP \\ \hline
P09 & Bachelor's & Black & 6 & Autism \\ \hline
P10 & Bachelor's & White & 7 & ADHD \\ \hline
P11 & High School & Black & 8 & GDD \\ \hline
P12 & Bachelor's & White & 5 & Neurodevelopmental Syndrome \\ \hline
P13 & High School & White & 16 & Motor \\ \hline
P14 & Master's & Asian & 13 & GDD \\ \hline
P15 & Master's & White & 9 & ADHD \\ \hline
P16 & Bachelor's & Asian & 11 & ASD \\ \hline
\end{tabular}
\end{table*}

All of our interview participants were women, and they all had children with DDDs. The youngest child in this group was two years old, whereas the oldest child was 17 years old. Each interview lasted for about an hour, and all participants received \$25 as remuneration for their contribution except one who wanted to volunteer in the interview study. The interviews were semi-structured. The underlying theme of our interviews was to understand the social support provided by YouTube videos on children's  DDDs. We followed the online social support theory proposed by Sheryl LaCoursiere \cite{lacoursiere2001theory} to formulate our interview questions. LaCoursiere defined online social support as the 1) cognitive, 2) perceptual, and 3) transactional process of initiating, participating in, and developing electronic interactions to seek beneficial outcomes in health care status, perceived health, or psycho-social processing ability. We formed our interview questions based on these three parameters and later pursued exciting topics raised by the participants. We asked them to describe when and why they started looking for YouTube videos on children's DDDs, how those videos impacted them to take necessary actions, how they verified the legibility of the information presented on YouTube, whether they encountered some misleading content, whether they interacted with other viewers of those videos through the comment section, whether they maintain those communications (if any) outside the YouTube platform, and whether they received emotional support as well as informational support.

 The interviews were analyzed using elements from grounded theory in particularly open qualitative coding \cite{khandkar2009open} using thematic analysis. After transcription, three researchers read through the interviews and developed themes individually based on the coded data material, which were later categorized. The themes emerged from the data rather than being prescribed from the interview guide. The researchers regularly met to iterate and converge on the broader categorization of themes. This analysis was performed to uncover parents’ experience in consuming YouTube videos on children's DDDs and how this content impacted them. The findings of the interviews are discussed in section \ref{fi}.

\section{Results: Styles and Content of Videos on Children's Developmental Delays}
\label{req1}
Our qualitative analysis identified four primary video styles (11 subcategories) in our dataset. Moreover, we found seven types of content in these videos. Female presenters presented 73.91\% videos, and 64.16\% presenters identified themselves as domain experts (such as doctors, researchers, therapists, EdTechs, counselors, and child psychologists). About channel topics, we identified eight unique channel topics for the videos in our dataset. However, little more than 75\% videos were created by ``education'' and ``people \& blog'' channels (45.37\% videos by education channels and 29.92\% videos by people \& blog channels). Table~\ref{tab2} lists all unique channels and the total number of videos generated by each channel type. Next, we briefly discuss the video styles and types of content identified through qualitative annotation.

\begin{table*}[!ht]
    \centering
    \caption{Topics of YouTube Videos in our dataset}
    \label{tab2}
    \begin{tabular}{|l|l|l|l|}
    \hline
        \textbf{Channel Topic} & \textbf{\# of Videos (\%)} & \textbf{Channel Topics} & \textbf{\# of Videos (\%)} \\ \hline
        Education & 696 (45.37\%) & Entertainment & 57 (3.77\%) \\ \hline
        People \& Blog & 459 (29.92\%) & News and Politics & 39 (2.61\%)\\ \hline
        Nonprofits and Activism & 138 (8.98\%) & Science and Technology & 34 (2.25\%) \\ \hline
        Howto \& Style & 78 (5.03\%) & Film and Animation & 31 (2.07\%)\\ \hline
    \end{tabular}
\end{table*}

\subsection{Video Styles}
Table~\ref{tab3} lists the four primary video styles we identified from qualitative coding. The table shows the sub-categories in each video style and the number of videos in each category. Example videos in each video style can be found in Figure \ref{fig2}.

The most frequently used video style was presentation slides. Videos in this category used slides to explain their topics. Some of the latest videos in this category included a small window to show the presenter. In contrast, the older videos in this category used only voice-over narration or subtitles imposed on the video.

The following frequent video style was presenting the topic using actors or characters. Videos in this category put effort into creating their videos in a natural setup. They often included experts in their videos who discussed critical topics and expressed their opinions on different example scenarios. News channels and morning shows created several videos in this category.

The third frequently found style was interview-based videos. Most of the videos in this category presented both the interviewer and interviewee in their videos, where interviewees were primarily experts in the domain. Some recent videos in this category showed Zoom setup for conducting interviews. They mostly talked about actions parents could take to care for their children with developmental delays when conventional therapy sessions were unavailable. A few videos in this category used podcast setup where the viewer could only hear the voice of the interviewer and interviewee.

The final category presented solo speakers. Parents often created these videos at home, sharing their experiences raising children with developmental delays. A few videos in this category were created by experts as well. Most videos in this category did not use any additional materials. However, experts often used graphs and flowcharts to clarify their points of view in this category.

\begin{table*}[]\small
\caption{Primary Video Styles \& Sub-categories}
\label{tab3}
\begin{tabular}{|
>{\columncolor[HTML]{FFFFFF}}l 
>{\columncolor[HTML]{FFFFFF}}l 
>{\columncolor[HTML]{FFFFFF}}l 
>{\columncolor[HTML]{FFFFFF}}l |}
\hline
\multicolumn{4}{|c|}{\cellcolor[HTML]{FFFFFF}\textbf{Video Styles}}                                                                                                                                                                                                                                                                                                                                                                                                                       \\ \hline
\multicolumn{1}{|l|}{\cellcolor[HTML]{FFFFFF}\textbf{Primary Categories}}                                                                     & \multicolumn{1}{l|}{\cellcolor[HTML]{FFFFFF}\textbf{Sub-categories}}                                                          & \multicolumn{1}{l|}{\cellcolor[HTML]{FFFFFF}\textbf{Count (\%)}} & \textbf{Description}                                                                                                                        \\ \hline
\multicolumn{1}{|l|}{\cellcolor[HTML]{FFFFFF}}                                                                                                & \multicolumn{1}{l|}{\cellcolor[HTML]{FFFFFF}\begin{tabular}[c]{@{}l@{}}Background \\ Narration\end{tabular}}                  & \multicolumn{1}{l|}{\cellcolor[HTML]{FFFFFF}7.54}           & \begin{tabular}[c]{@{}l@{}}Narrator explained the actions \\ of the characters/actors in the \\ background of the video\end{tabular}        \\ \cline{2-4} 
\multicolumn{1}{|l|}{\cellcolor[HTML]{FFFFFF}}                                                                                                & \multicolumn{1}{l|}{\cellcolor[HTML]{FFFFFF}Expert opinions}                                                                  & \multicolumn{1}{l|}{\cellcolor[HTML]{FFFFFF}12.39}          & \begin{tabular}[c]{@{}l@{}}With narrator's explanation, \\ experts were introduced in \\ these videos for short bites\end{tabular}          \\ \cline{2-4} 
\multicolumn{1}{|l|}{\multirow{-3}{*}{\cellcolor[HTML]{FFFFFF}\begin{tabular}[c]{@{}l@{}}Videos with \\ actors/characters\end{tabular}}}      & \multicolumn{1}{l|}{\cellcolor[HTML]{FFFFFF}Subtitles}                                                                        & \multicolumn{1}{l|}{\cellcolor[HTML]{FFFFFF}4.07}           & \begin{tabular}[c]{@{}l@{}}No narration or expert bites\\  were used, all explanations \\ presented through subtitles\end{tabular}          \\ \hline
\multicolumn{1}{|l|}{\cellcolor[HTML]{FFFFFF}}                                                                                                & \multicolumn{1}{l|}{\cellcolor[HTML]{FFFFFF}\begin{tabular}[c]{@{}l@{}}With presenter's \\ video and voice\end{tabular}}      & \multicolumn{1}{l|}{\cellcolor[HTML]{FFFFFF}9.81}           & \begin{tabular}[c]{@{}l@{}}Presenter's face appeared \\ on the side of the slides along \\ with their narration\end{tabular}                \\ \cline{2-4} 
\multicolumn{1}{|l|}{\cellcolor[HTML]{FFFFFF}}                                                                                                & \multicolumn{1}{l|}{\cellcolor[HTML]{FFFFFF}\begin{tabular}[c]{@{}l@{}}without presenter's \\ video, only voice\end{tabular}} & \multicolumn{1}{l|}{\cellcolor[HTML]{FFFFFF}22.25}          & \begin{tabular}[c]{@{}l@{}}No face appeared, only \\ narration was included\end{tabular}                                                    \\ \cline{2-4} 
\multicolumn{1}{|l|}{\multirow{-3}{*}{\cellcolor[HTML]{FFFFFF}\begin{tabular}[c]{@{}l@{}}Videos made of \\ presentation slides\end{tabular}}} & \multicolumn{1}{l|}{\cellcolor[HTML]{FFFFFF}Only subtitles}                                                                   & \multicolumn{1}{l|}{\cellcolor[HTML]{FFFFFF}6.87}           & \begin{tabular}[c]{@{}l@{}}No presenter's view or narration \\ were included, only subtitles \\ included with background music\end{tabular} \\ \hline
\multicolumn{1}{|l|}{\cellcolor[HTML]{FFFFFF}}                                                                                                & \multicolumn{1}{l|}{\cellcolor[HTML]{FFFFFF}\begin{tabular}[c]{@{}l@{}}Face-to-Face \\ Interviews\end{tabular}}               & \multicolumn{1}{l|}{\cellcolor[HTML]{FFFFFF}13.15}          & \begin{tabular}[c]{@{}l@{}}Both interviewer and \\ interviewee appeared in the \\ same location for the interview\end{tabular}              \\ \cline{2-4} 
\multicolumn{1}{|l|}{\cellcolor[HTML]{FFFFFF}}                                                                                                & \multicolumn{1}{l|}{\cellcolor[HTML]{FFFFFF}Zoom interviews}                                                                  & \multicolumn{1}{l|}{\cellcolor[HTML]{FFFFFF}5.72}           & Zoom setup was used for interviews                                                                                                          \\ \cline{2-4} 
\multicolumn{1}{|l|}{\multirow{-3}{*}{\cellcolor[HTML]{FFFFFF}Interview videos}}                                                              & \multicolumn{1}{l|}{\cellcolor[HTML]{FFFFFF}Podcast interviews}                                                               & \multicolumn{1}{l|}{\cellcolor[HTML]{FFFFFF}3.69}           & \begin{tabular}[c]{@{}l@{}}Static screen was used along with \\ the voice of the interviewer and \\ interviewee taking turns\end{tabular}   \\ \hline
\multicolumn{1}{|l|}{\cellcolor[HTML]{FFFFFF}}                                                                                                & \multicolumn{1}{l|}{\cellcolor[HTML]{FFFFFF}\begin{tabular}[c]{@{}l@{}}With video clips, \\ images, slides\end{tabular}}      & \multicolumn{1}{l|}{\cellcolor[HTML]{FFFFFF}5.45}           & \begin{tabular}[c]{@{}l@{}}Solo speaker intermittently used clips,\\  images, graphs in support of their opinion\end{tabular}               \\ \cline{2-4} 
\multicolumn{1}{|l|}{\multirow{-2}{*}{\cellcolor[HTML]{FFFFFF}\begin{tabular}[c]{@{}l@{}}Videos with \\ solo speakers\end{tabular}}}          & \multicolumn{1}{l|}{\cellcolor[HTML]{FFFFFF}\begin{tabular}[c]{@{}l@{}}Without any \\ supporting materials\end{tabular}}      & \multicolumn{1}{l|}{\cellcolor[HTML]{FFFFFF}9.06}           & \begin{tabular}[c]{@{}l@{}}No supporting materials used \\ in the solo presentation\end{tabular}                                            \\ \hline
\end{tabular}
\end{table*}

\subsection{Types of Video Content}
We identified seven types of content in our dataset. Two types of videos were primarily created by parents of children with developmental delays (specified as ``primarily created by parents'')—more than 72\%  of videos covered more than one type of content. Thus, we counted those videos more than once to calculate percentages. 

\subsubsection{Defining Concepts}
The first category of videos in our dataset defined various concepts related to children's developmental delays. Professional experts such as researchers, pediatricians, and therapists primarily created these videos to explain different topics, such as delays in cognitive development, motor development, social and behavioral development, and physical development. Videos in this category often used text, flowcharts, graphs, and animation to explain these concepts.

\subsubsection{Explaining Signs of Developmental Delays}
Videos in this category play a crucial role in explaining the early symptoms parents can monitor to examine if their child has any developmental delays. Most videos in this category used short clips of children's interaction to present their topic. One major concern regarding DDDs is to identify the underlying medical condition based on various symptoms. Young children often show similar symptoms for a wide range of DDDs. Experts in their videos highlight the subtle differences in those symptoms to help parents distinguish between developmental delays and disabilities and their subcategories. Moreover, they use YouTube as a widely accessible medium to warn parents that some diagnoses cannot be done at home and should only be done by experts at clinical facilities. Thus, parents should consult their pediatricians or other developmental experts for those evaluations instead of deliberating, underlining the importance of early detection and professional consultation.

\subsubsection{Presenting treatments to Developmental Delays}
Physicians and therapists, in their videos, discussed various facilities where children with developmental delays can receive consultations, necessary treatments, and training. We observed that at least 35\% of videos in this category were created by television channels such as BBC, ABC News, and CBS. These videos presented a short documentary about a facility that provided training, counseling, and necessary guidance to children with developmental delays, emphasizing the wide range of resources and support available for these children and their families.

\begin{figure*}
\begin{center}
\includegraphics[scale=0.5]{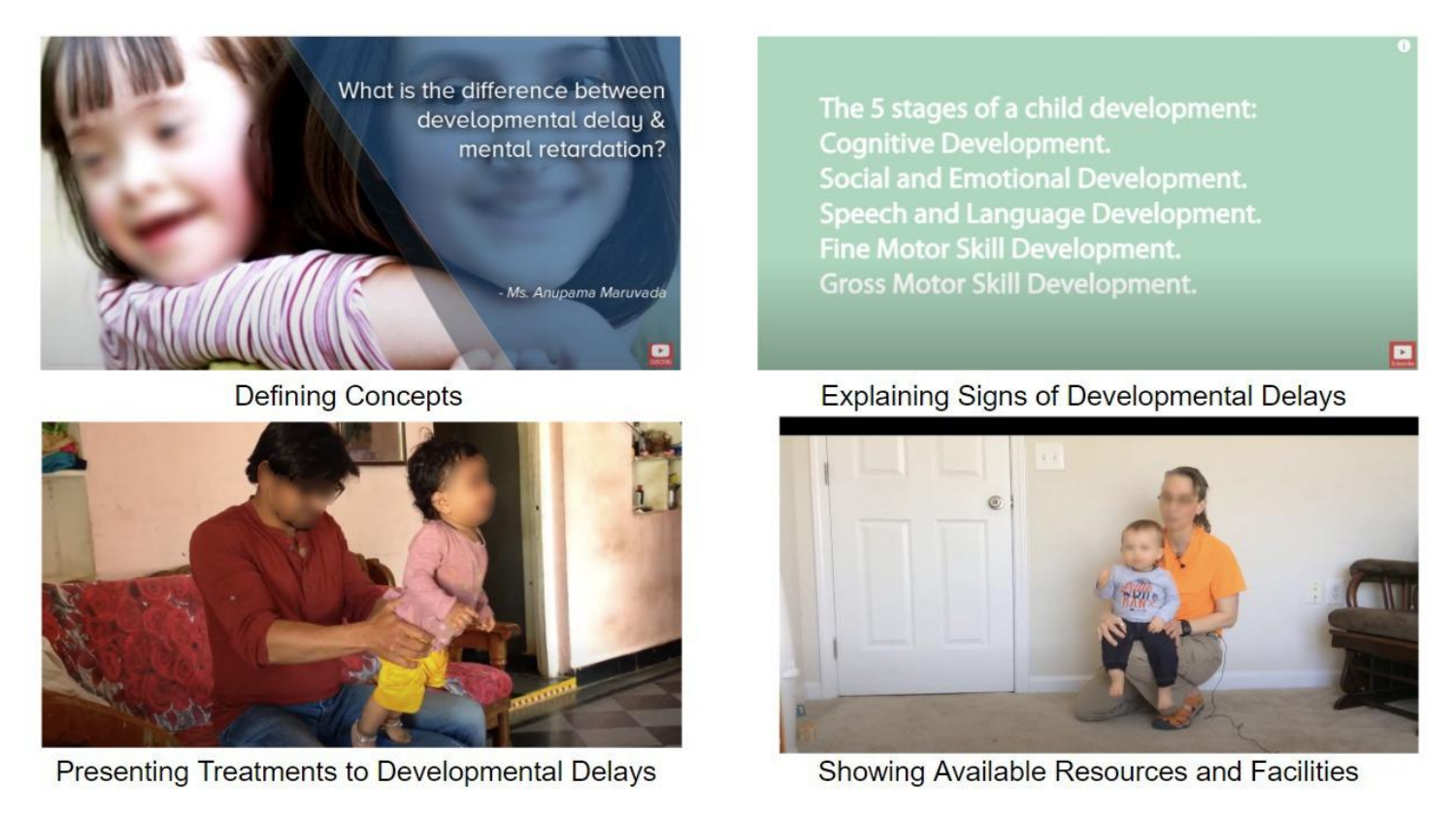}
\includegraphics[scale=0.5]{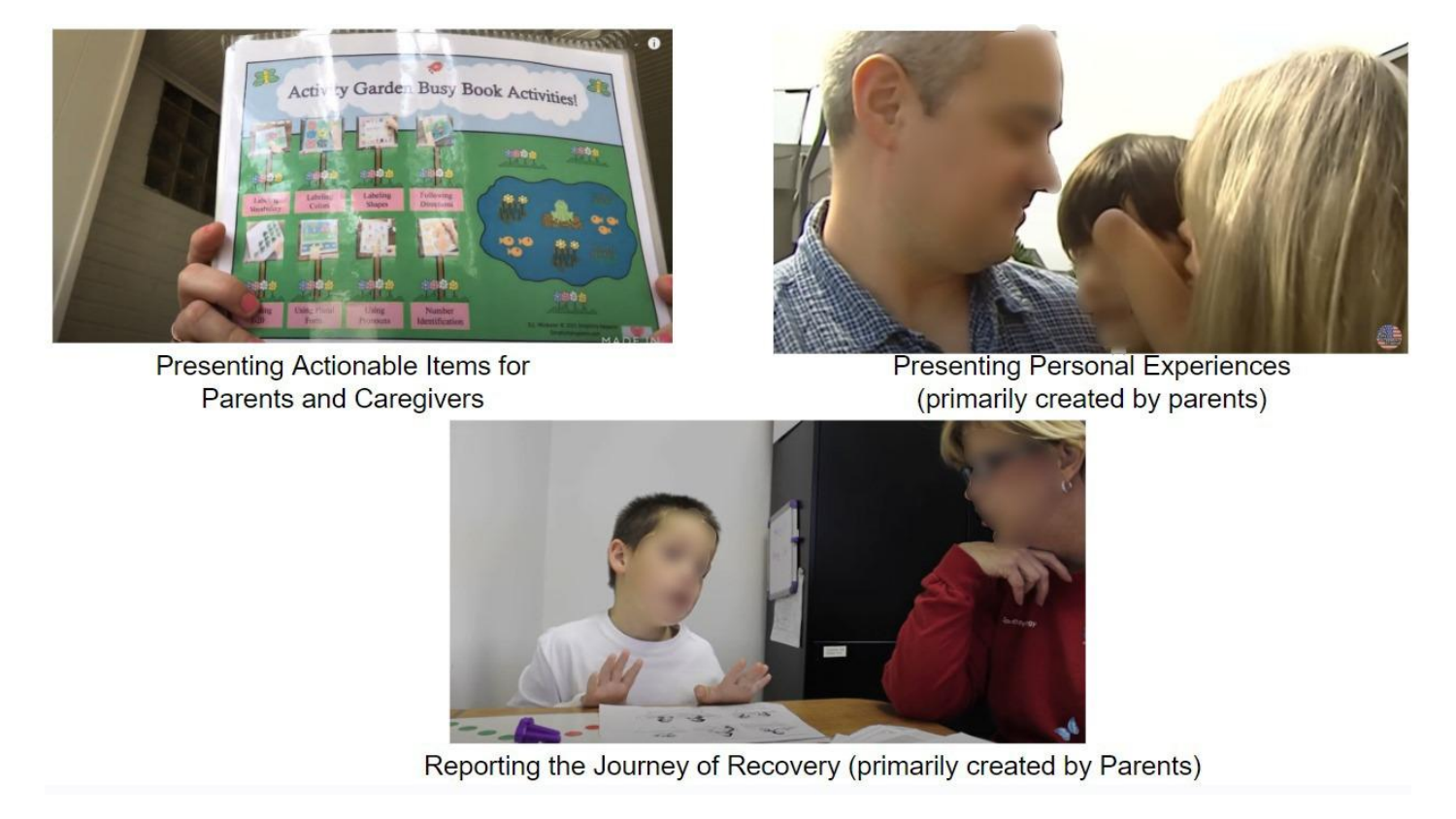}\\
\caption{Types of Video Content identified from our dataset}
\label{fig1}
\end{center}
\end{figure*}

\subsubsection{Showing Available Resources and Facilities}
Videos in this category presented a range of services and facilities available for children with developmental delays instead of focusing on one specific facility. Presenters often focused on local facilities in these videos and explained how parents could enroll or nominate their children for these facilities. Presenters also showed alternative resources if parents needed help enrolling their children in state-sponsored programs.

\subsubsection{Presenting Actionable Items for Parents and Caregivers}
Videos in this category talked about steps and routines parents can follow at home if their children are diagnosed with a specific developmental delay. For treating children with developmental delays, most of the time, parents must invest a significant amount of time and energy and adhere to the practices and routines that their counselors suggest. These videos discuss how parents can incorporate these routines into their daily schedules. In addition, they present apps (GEMINI), books (105 Activities for Your Child With Autism and Special Needs: Enable them to Thrive, Interact, Develop and Play), toys (What Goes Together? Activity Box), and gadgets that parents can consult if they need guidance in this regard. Experts in these videos also explained the mental and emotional challenges that children go through, which they can hardly express to their parents and caregivers. Experts in their videos provided suggestions to parents and caregivers on how to handle those conditions (such as children with cerebral palsy in a crowded airport) in a way so that their children with DDDs feel secure, confident, and comfortable in difficult situations.

\begin{figure*}
\begin{center}
\includegraphics[scale=0.5]{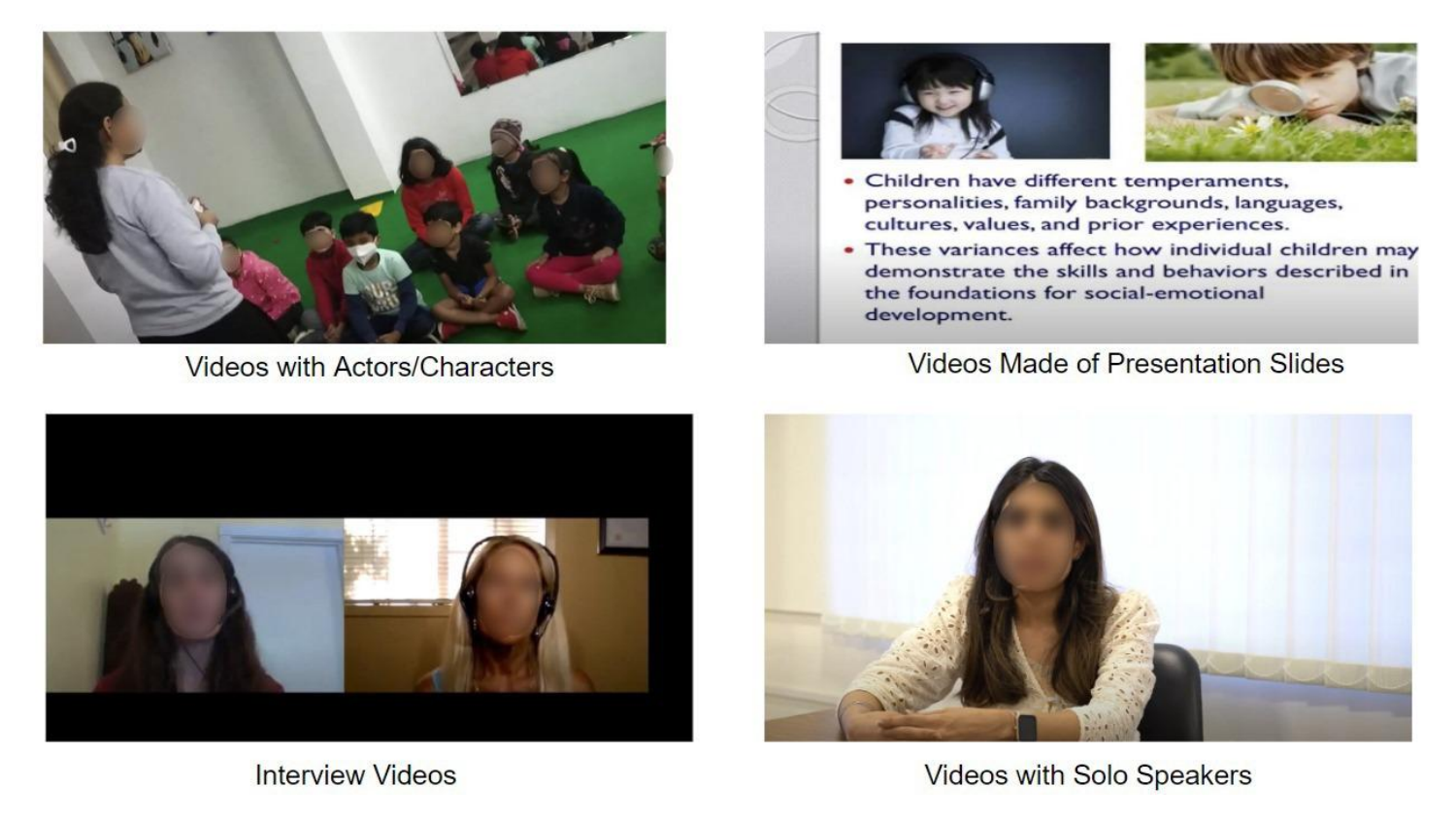}\\
\caption{Primary Video Styles}
\label{fig2}
\end{center}
\end{figure*}

\subsubsection{Presenting Personal Experiences (primarily created by parents)}
Parents of children with developmental delays primarily created videos in this category. In these videos, parents explained their journey and the challenges of diagnosing developmental delays among their children. They discussed how it impacted their married life, parenting practices with other kids, professional commitments, and social life. Parents also discussed how other parents could consider taking action proactively to avoid uncertainty and mental stress.

\subsubsection{Reporting the Journey of Recovery (primarily created by parents)}
This is another category of videos that parents created. Parents often made follow-up videos several months after their initial video to discuss the progress of their children diagnosed with developmental delays. They explained the services and treatments they received for their children. They also explained the merits and demerits of those services that they received. In these videos, they often introduce their children and show short moments of interaction with them. As parents mentioned, the primary objective of making these videos was to assure other parents that developmental delays among young children are not uncommon and that, with proper care, children can overcome their delays entirely and perform like other children with early delays. On the other hand, parents of children with non-curable conditions such as ADHD and Cerebral Palsy appeared on YouTube to educate others that no therapies or interventions can fully cure such conditions. However, those interventions helped their children manage their symptoms and improved their lifestyle.  

\subsection{Summary}
In our analysis, we identified four video styles primarily. In addition, we found seven types of content in our dataset. This analysis helped us understand both experts and parents of children with DDDs create video content on YouTube. The expert videos focus on the informational aspect of DDDs. In contrast, the parent videos highlight the personal challenges in a wide range of scenarios that other parents face while raising their child with DDDs and in some cases, how they overcome them. The findings in this section will help us understand the regular interactions and topics of discussion of the audience in the comment section. In addition, these findings will help us understand the broader impact of these DDDs videos on parents of children with DDDs.   

\section{Results: User Involvement examined from YouTube Comments}
\label{sci}
In RQ2, we aimed to explore how these impact the parents and caregivers of children with DDDs. To this end, we took two approaches: 1) analyzing the comments of the videos in our dataset and 2) interviewing parents of children with DDDs who watched videos on YouTube on children's DDDs. In this section, we presented the main topics that parents and caregivers discussed in the comment section. In the next section (section~\ref{fi}), we will show the observations made by qualitatively analyzing the interview transcripts.

From our dataset of 1532 videos, we excluded 121 videos as the video creators blocked their comment property. The remaining 1411 videos were used in our comment analysis. We performed qualitative coding and identified six primary themes for comments. These themes are discussed here.

\subsection{Asking for suggestions}
This was the most common category of comments. Parents explained the condition of their children related to the topic of the video and asked for suggestions from experts and other fellow parents. Often, they directly asked video creators whether they could help their children. In many comments, parents mentioned that the video helped them understand the milestones of a specific type of development. The information presented in the video helped them realize that their children might have some form of DDDs, for which they asked for expert suggestions. As a community, parents asked detailed queries about services supported by health insurance, which developmental specialists often recommend. Other parents helped each other by giving suggestions that might help parents negotiate better with insurance companies or use treatment codes that are more likely to be reimbursed. 

\subsection{Sharing Experiences and Concerns}
This was the second most frequent type of comment in our dataset. Parents used the comment section to report the status of their children's DDDs. They discussed the treatments their children received for their delays and how well they responded to those treatments. They also described different aspects of treatment programs, such as the lack of facilities in suburban areas, the high cost of specialized therapies, and the challenges of continuing the treatment program for a long duration. Parents also described how they followed all suggestions in similar videos, but their children did not respond. We observed that parents sought a sense of support from the community in their comments when they just became aware of their children's DDDs but did not yet know how to go through the treatment, therapy, and counseling programs.

\begin{table*}[!ht]
    \centering
    \caption{Examples of Six Comment Themes}
    \label{tab4}
    \begin{tabular}{ | m{12em} | m{9cm}| }
    \hline
        \textbf{Category of Comments} & \textbf{Example of Comments} \\ \hline
        Asking for suggestions (38.51\%) & My child just turned 3 and he hardly likes to play with other kids, he easily gels up with adults but with kids he seems little nervous or scared can u suggest some more things which i can do to improve his this skills. \\ \hline
        Sharing Experiences and Concerns (23.16\%) & My baby is 1 year and 2 months...he haven't been able to stand by his own and neither to walk...he walks only with support....iam worried now... \\ \hline
        Thanking Video creators and Asking for New Content (11.63\%) & Hello dear. I really like your videos and took help from them, so thanks for that, Can you please make a video on problem solving component as u discussed there... \\ \hline
        Response of Video Creators (14.69\%) & Thank you for taking the first step and expressing your struggle. We want you to know you are not alone in how you feel. We hope you will check out the links to the resources in the description box for where to learn more and where to get treatment and support.\\ \hline
        Comments from Fellow Video Creators (4.14\%) & Hi, This is a very useful video. I'm working on a project related to women and children health among immigrants in Portugal. Specially with early detection and intervention. Would like to discuss with you about making an awareness video for the beneficiaries.\\ \hline
        Miscellaneous (7.87\%) & Doc Oyalo can reverse autism with herbs and it’s completely perfect. I used it for my son and so far his speech is verbal and social skill is normal and he can now also respond to everything positively on his own\\ \hline
    \end{tabular}
\end{table*}

\subsection{Thanking Video Creators and Asking for New Content}
This category includes a list of comments in which the audience thanked video creators for sharing content on YouTube. Parents thanked video creators for information they needed help finding through traditional web searches. In these comments, parents frequently requested video creators for more content on similar topics. They also asked for further explanations of the topic presented in the video in this category of comments. In thank you notes, parents often referred to other videos of the same creator or other similar creators that helped them take necessary actions for their children's DDDs. 

\subsection{Responses and Reflections of Video Creators}
Video creators used the comment section to communicate actively with their audience. They acknowledged their viewers' thank you notes and reassured their audience about preparing new content in their comments. They also responded to their audiences' queries and sometimes included information for additional resources in their comments. For instance, in one comment, the video creator explained that as they are not certified to diagnose DDDs among children, they could not tell whether a child has a developmental delay. Instead, they request parents consult a certified physician for that task.

Video creators sometimes focus on reacting to the audience's comments. Instead, sometimes, they proactively asked their audience about the progress of their children's treatment and therapy sessions. Video creators use the comments section of their videos to create personal connections and a strong bond with their audience. Many of these creators used the comment section to talk about their connection with children of DDDs and how those experiences changed their point of view on this issue.

\subsection{Comments from Fellow Video Creators}
Video creators received comments from parents of children with DDDs and fellow experts and activists in their domain. Fellow experts appreciated video creators for creating content on DDDs. They highlighted critical elements of the videos and explained how those content would help parents to make early decisions for their children. Fellow experts often suggested video creators for adding additional content to their videos to make them easier to understand for layman users with little to no knowledge about DDDs. Fellow experts also tried to communicate with video creators for future collaboration opportunities. They often invited video creators to collaborate and create new videos on YouTube. Apart from those invitations, they also offered the video creators the opportunity to join offline collaborations such as voluntary services in community clinics. It is essential to mention that fellow experts only sometimes praise video creators in their comments. Sometimes, they claimed that the video's content needed to be corrected and suggested that video creators change their content to avoid confusion.

\subsection{Miscellaneous}
Around 8\% comments did not fit into any of the five categories discussed above. We included them in this ``miscellaneous'' category. One group of comments in this category addressed the effectiveness of alternative treatments for DDDs. For instance, several comments discussed the impact of herbal therapy offered by Dr. Oyalo that completely cured speech delay and autism for young children. Other viewers also ridiculed such claims in their comments as they identified Dr. Oyalo as a ``modern-day snake oil salesman''. Another group of comments in this category discussed unrealistic expectations of family's close relatives regarding children's developmental milestones. For instance, one mother complained about her in-laws, who expected their three-month-old granddaughter to be able to sit independently. When the child could not meet their unrealistic expectations, her in-laws started complaining that the child had some growth delay. Finally, a few comments in this category discussed the rapid progress of their over-achiever kids. For example, in one such comment, parents asked whether they should worry about their child who reached all developmental milestones recommended for 5-year-old kids at the age of 2.5 years.

\subsection{Summary} 
Overall, the analysis of the comments showed us how the community (video creators, parents of children with developmental delays, physicians, therapists, counselors, and researchers) used the comment section of YouTube to build a strong bond among each other. Parents used this platform to search for and learn critical information on developmental milestones, disabilities, and delays. They also interacted with community members to clarify doubts and find necessary resources. The comment section allowed parents to participate actively in community-wide conversations, inspiring video creators to stay engaged with the community and produce more content demanded by the audience. Experts used this platform to interact directly with their audience and fellow video creators, creating a sense of social bond with the broader community.

\section{Findings of the Interview of Parents and Caregivers (RQ2)}
\label{fi}

\subsection{Accessible Source of Information}
All participants emphasized the importance of YouTube videos as the first point of information on DDDs. All but three participants mentioned that their children were not diagnosed with any form of DDDs at the time of birth. Instead, they gradually started noticing growth and behavioral patterns that were not usual over several months. However, these observations were insufficient for them to raise concerns with their pediatricians. They all experienced a sense of guilt, denial, and a lack of confidence to admit their observations. In some scenarios, their spouse was not ready to accept any form of DDDs for their children(N = 2). Such denials delayed the diagnosis for one child more than two years. YouTube videos on DDDs worked for them as a first point of information for which they did not have to reach out to external resources. The videos were free, easily accessible, and, to a large extent, easy for novice users to understand. The availability of these videos allowed them to explore in their comfort instead of asking for suggestions from friends and family members, who often become judgmental in similar cases. As one mother mentioned: 

"\textit{We were with my husband in his European duty station. Our pediatrician never asked us to fill out any developmental surveys. But there was something that bothered me all the time. Luckily, one day, I found a channel on YouTube where a developmental specialist was talking about early signs of developmental disabilities. The information shared on those videos finally convinced my husband that our daughter needed to see some specialists. It turned out that our daughter had GDD (Global Development Delay) because of some abnormalities in the corpus callosum and a mutation on ARF1.}"[P7]

The YouTube Videos did not only assist parents during the initial stage of clearing doubts but also worked as a consistent source of information after the diagnosis. As P11 mentioned:

"\textit{The diagnosis was only the beginning of our journey. Taking care of a child with a developmental disability requires continuous searching for the best therapy centers, the best schooling, and the best places to socialize with other children. Only expect to receive some of that information from our pediatricians. We only meet her once every three months unless there is an emergency. Parents like us received much useful information regarding occupational therapy, feeding therapy, or Medek therapy by following YouTube channels on DDDs.}"[P11]

Some other parents (P4, P9, P12, P14) explained YouTube as a complementary social media platform to Facebook. On many occasions, mothers of children with DDDs form private Facebook groups to discuss their queries and concerns. P4 remembered multiple incidents when someone in the Facebook group suggested others watch informational videos on YouTube in response to some queries. Moreover, some parents consider YouTube an excellent resource for verifying information they receive from Facebook groups. 

"\textit{[...] I learned from my Facebook group about a center close to my home that provides social skills therapy. I immediately searched for that center on YouTube and found their promotional video. The video gave me a rough idea about the look and feel of the center, and I decided to visit the place in person. As a parent of an autistic child, I cannot afford to visit all potential centers in person. Such promotional videos or virtual tours give me critical information about whether someplace is worth visiting.}" [P9]

Participants found that videos featuring clinical specialists and therapists were rich with information. Those videos helped them understand their children's condition better. However, they found that videos made by other parents or caregivers were equally important as those videos often provided helpful tips on supporting their children emotionally. As P2 mentioned:

"\textit{The physical and occupational therapies are necessary for my child to grow and learn essential life skills. But therapists do not stay with my child 24X7. Therapy sessions did not tell me how to deal with my son's emotional meltdowns. Videos made by other parents helped me realize how to handle those difficult moments and support my son through those phases of overwhelming experiences of emotions.}"[P2]

Some parents (N=4) said consuming information is relatively easier on YouTube than other web search engines. As P6 mentioned:

"\textit{[...] My daughter has a rare condition which is not even known to most of the pediatricians. Ideally, I would like to read all research papers published on that condition so that I can provide the best care to my daughter. However, realistically, I feel exhausted at the end of the day and generally have no energy to read a medical journal. Yet, I can watch YouTube even when I am tired, and if I find an interesting video the next morning, I find the original publication and read it thoroughly. So, in a way, YouTube's video format makes the research on TBCD disorder more manageable and accessible for me.}"

\subsection{Experiencing a Sense of Virtual Support}

All participants expressed the importance of forming a passive online community not only through the comments section of YouTube videos but also from watching videos that other parents of children with DDDs made. Some participants (N = 9) discussed their inability to be part of regular parent meetings or family get-togethers because of the lack of social acceptance for children with DDDs. This condition made them feel isolated and lonely, which made them more vulnerable to depression. YouTube videos helped them feel a sense of social presence. These videos enabled them to find in-person and virtual communities where they were accepted without judgment. More importantly, these videos gave them the courage and strength to deal with their daily challenges, starting from training their children to go to the toilet on their own at the age of 10 (regularly developed children typically get potty trained at three years of age) to teach them how to talk or behave in a socially acceptable way with other children.

Some parents (N = 3) shared their experiences of creating personal connections with other parents whom they initially started communicating with through YouTube's comment section. Unlike some other entertaining YouTube channels where viewers frequently attack each other at a personal level, viewers of DDDs-related YouTube channels always showed a sense of empathy and a caring attitude to others. As P12 described:

"\textit{Initially, when my son was diagnosed with neurodevelopmental syndrome, I constantly fought with my husband. I am not saying that it was his fault or mine. I guess we both felt frustrated and defeated. It was just like the book "Welcome to Holland" where you got on the flight to go to Italy but somehow ended up in Holland and the sense of getting stuck there for the rest of your life. Other parents' experiences (shared on YouTube) gave us the sense of belief that with resilience and patience, we can do this too.}"[P12]

Other participants (N = 2) also talked about their relationship with their regularly developed children. Many videos created by parents on YouTube not only include their child with DDDs but also talk about their struggle to allocate sufficient time to their other regularly developed child. As P11 has described, family dynamics are different. So, she expected to avoid mimicking the strategies that worked for other parents. Instead, she found strength and motivation from other's videos and started experiencing a renewed sense of belief in her own family's strength.

\subsection{Experiencing a sense of seclusion and recluse}

Social media platforms are widely regarded as convenient places for creating virtual support communities and collaborative groups for people with specific needs. However, unlike other social media platforms, some participants (N = 6) considered YouTube a perfect blend of seclusion and connection. As P1 mentioned:

"\textit{No matter how much support you get from hundreds of resources, this journey is painful. You cannot ignore it. My son goes through good phases as well as bad phases. During his bad phases, I want to stay alone. I wouldn't say I like to meet other parents in the playgroup, but I do not want to talk to people on Facebook. YouTube gives me that sense of seclusion without cutting me off from this virtual world.}" [P1]

In other cases, parents also discussed the anonymity features of their YouTube account compared to their public account on Facebook. On YouTube, parents can choose to use pseudonyms instead of using their real names. This feature allows them to maintain a level of privacy when they are discussing their child's developmental challenges. As P11 mentioned:

"I sometimes hesitate to share sensitive details about my daughter on Facebook. You know, some of the Facebook groups for GDD (Global Developmental Delay) and ADHD are not private. When I post some of my concerns on Facebook, they could be exposed to all my friends, family members, and even my colleagues at work. That is different from what I want, at least not always. YouTube gives me that subtle sense of separation, but at the same time gives me access to a community to reach out to at the time of need."

Some participants (N = 5) also discussed their emotional involvement in social media groups with other parents. When browsing groups on various social media platforms, they often find many personal stories. Many of them are emotionally draining, and they could relate to many of those shared experiences with their own experiences. Participants usually avoided those emotional burdens and consumed only informational content. As P7 mentioned:

"\textit{[...] It is not easy to accept that your child cannot do simple tasks alone. It may sound not good, but I need a break as well. Facebook groups are full of such stories. Sometimes, I do not open them for days to recollect myself. Sometimes, I watch YouTube videos to learn about new research findings during this time, but I do not read the comment section.}" [P7]

\subsection{Verifying the Credibility of Information}

All participants discussed their points of view on the credibility of information presented on YouTube about DDDs. The spread of misinformation and disinformation during the COVID-19 pandemic made parents more vigilant when consuming health-related information. All participants highlighted the importance of checking the credentials of the video creators on YouTube. As P2 mentioned:

"\textit{YouTube is a warehouse of knowledge. Anything you name, you will find it on YouTube. But too much information also comes with too much noise. Sometimes, it is hard to differentiate between a credible one and a non-credible one. That is why I always rely on videos that clinical specialists create. I verify their credentials using other channels, such as Google's search engine. It is hard to fake in two different places.}"

Other parents (N = 8) discussed following only channels created by reputed organizations such as the World Health Organization (WHO). This strategy is comparatively safe; however, participants agreed that they could only sometimes follow this rule as one organization, even the large ones, cannot produce sufficient content. In those cases, participants often relied on repeated occurrences of information. When parents found information in multiple videos produced by different channels, they started trusting that information more.

Some participants considered YouTube and other social media platforms as only the initial source of information on DDDs. If they find something useful, they always consult it with their pediatricians before adopting them in any capacity. As P4 mentioned: 

"\textit{No matter what precautions you take, you can never be sure of the authenticity of information found on YouTube. One may feel the temptation to try them right away. [...] I cannot take chances on my kid's healthcare. I always double-check such information with my pediatricians and behavioral therapists before trying them on my child. }"

\subsection{Summary: Interview Study}

These interviews showed us the breadth and depth of interactions that parents of children with DDDs experience while consuming videos on YouTube. YouTube provides opportunities for parents not only to consume content but also to build a solid virtual community. Our analysis revealed the critical need for seclusion in the age of social media, especially for those populations with highly demanding responsibilities. The findings also helped us reflect on how parents consume YouTube videos on DDDs of different video styles and types of content and how those content types impact parents' trust.   

\section{Discussion and Design Implications}
\subsection{YouTube as a Source of Informational Support}
Developmental delays and disabilities among young children may bring a wide range of challenges for parents, especially when parents have no specific reasons to predict such conditions. Observing the early signs of DDDs among young children is not an easy task. It often takes a trained pair of eyes. Yet, many parents do not have the provisions to consult experts at the right time because of their socio-economic conditions. For others, there are high chances of missing the subtle signs and symptoms that only experts can distinguish without confusion. YouTube videos on DDD are a great source of information for parents.

Our analysis showed that parents considered YouTube as an accessible search engine for information on DDDs. YouTube offers unrestricted access to videos and maintains users' privacy by not displaying subscribers' lists, thus providing privacy with content access. Parents especially appreciate YouTube for seeking practical and actionable information to support their children's developmental growth from early stages. As we observed in our interviews, some parents overcome their state of denial by passively learning about DDDs from YouTube. Others use them from the urge to advocate for their children with their primary care pediatricians. These informational videos give parents the confidence to ask questions about their children's overall well-being and developmental growth, sometimes overlooked during regular wellness visits.
Additionally, parents and caregivers favored longer videos over shorter reels for informational content on children's DDDs and the personal-experience stories from other parents, which provided them with a sense of connection and assurance. Although parents found valuable information from various social platforms, they frequently visited YouTube for more detailed insights. The platform's how-to videos and personal testimonies were more relatable for understanding their child's overall developmental growth. 

Here, it is critical to understand that the information provided on YouTube, even by experts, cannot be processed as medical advice. Previous work reported that the quality of medical information on YouTube could be of better quality in general \cite{osman2022youtube}. Researchers also observed that YouTube may promote misinformation, especially regarding highly controversial topics (such as vaccination). Misinformation on sensitive issues such as children's DDDs can have long-term and deadly consequences. Channels promoting videos on DDDs need to ensure that the quality of the information is not compromised. As a platform, YouTube can play an influential role in this context by rewarding reliable, expert-driven channels with unique tags for generating certified content on medically critical domains. This will give parents and caregivers more agency in what they learn about children's DDDs on YouTube.

In our work, we investigated the video styles and the content of videos on DDDs. The qualitative analysis revealed how experts use YouTube to explain complex, hard-to-discern ideas more understandably so that parents and caregivers from all backgrounds can consume them and take necessary actions as appropriate. Although the scope of this paper considered all DDD-related videos in a single group for analysis, they can also be sub-categorized into smaller granularities. For example, developmental delay-related videos can be further classified for various delays, such as cognitive, motor skills, speech, and social and behavioral. Developmental delays and disabilities can be divided into separate lists for further classifications. Here, it is worth mentioning that although developmental delays vs. disabilities have significantly different trajectories for adults and children during their initial years, the differences are not always clearly observable. Moreover, children in many scenarios experience multiple types of developmental conditions. Future work may investigate how parents find such classifications valuable and convenient for quick and efficient access to these videos.    

\subsection{A Mutually Supportive Online Community for Emotional Assistance}
We observed YouTube video content and the discussion community as a constant source of emotional support system for parents and caregivers of children with DDDs. The personal experiences shared by other parents provided a sense of hope and motivation and helped them feel part of a community rather than isolated. Our analysis resonated with previous work that discussed emotional support from social media platforms~\cite{shensa2016social, hu2022understanding}, whereas some of our observations added new knowledge in this domain. For example, we found the need for seclusion without completely withdrawing from social media platforms. 

On sensitive issues, emotional support can be overwhelming as well, depending on the mental state of the user. One such example can be the family experiences shared by other parents. While in the majority of the scenarios, personal experience-sharing videos are created in an empathizing way, parents may find them emotionally straining in certain conditions. As P7 discussed about her son being at the higher end of the autism spectrum, experiences shared by other parents with the lower end of the autism spectrum did not always match with her own experiences. Stories of gradual progress and improvement may come at different paces for different parents, irrespective of their personal efforts, treatments, and interventions. Emotionally relevant stories may become hard to process during those difficult times. We have seen in prior work similar concerns where Long-COVID patients felt a sense of guilt while sharing their recovery stories~\cite{karra2023fishing}. Platforms like YouTube should consider having provisions where users can control their YouTube suggestions. This control will allow them to receive no suggestions for personal experience sharing videos for a specific period, allowing them to recover from their state of mental stress. This design can also be extended to the comments section of the videos, where users may have the control not to see any comments for the videos they watch. Gaining more control over the content of the videos that parents and caregivers prefer to watch and how they want to watch them will be beneficial for the broader community, especially inclined to consume health-related information from YouTube.

\subsection{Authenticity of Health Information on YouTube}
Sharing healthcare information on social media often receives much-concerned attention from physicians and policymakers. The lack of information control on social media platforms like YouTube bears the risk of spreading misinformation and rumors, which can seriously affect public healthcare~\cite{tasnim2020impact, dhengre2022impact}. Despite these concerns, previous research established niche areas in healthcare where YouTube can have a significant positive impact~\cite{sharif2021positive}. Our work contributes to that line of research. In addition, we acknowledge the importance of verifying the validity of critical healthcare information on social media. One way to address this concern is to provide supporting information from verified resources (such as CDC, Mayo Clinic, and Johns Hopkins Pediatric Care) attached to informational videos on YouTube about children's DDDs. Experts creating informational videos on YouTube may promote this practice of providing authentic sources of information enclosed as an attachment to their videos. As a platform, YouTube may encourage this initiative by algorithmically upvoting these videos to appear at the top of the search query. The platform can also create curated playlists focused on such verified videos and arrange them sequentially from early intervention to advanced treatment management strategies for various developmental delays and disabilities. These playlists can guide viewers through a comprehensive learning journey, providing reliable, relevant, and actionable information at every stage of their child's developmental growth.

\section{Conclusion and Future Work}
DDD is a condition that is frequently observed among young children. Early interventions, in most cases, can help children recover completely from their childhood delays or improve partially overcoming their disabilities. Unfortunately, most children cannot access early intervention programs. Studies found that even when facilities were available to care for children with DDDs, children were not recommended for those services due to a lack of awareness. Thus, parents of children with DDDs went through parental stress, which caused social isolation. YouTubers have created thousands of videos on children's DDDs, both from the perspective of experts and parents raising delayed children. We conducted a mixed-method analysis of YouTube videos on DDDs to identify the primary content and style of videos about DDDs. We found informational videos created by experts and emotionally connecting content created by parents of children with DDDs. Our analysis of the comments in these videos showed how YouTubers and viewers created a strong connection through these videos.
Most importantly, the interview study shows how these videos may help parents not only gain information about CDs but also overcome various emotional challenges. This work is the first step in building a community on children's DDDs. The findings of our work can help develop reliable and relatable content storage for children's DDDs and can be expanded for other healthcare communities as well.

\bibliographystyle{ACM-Reference-Format}
\bibliography{sample-sigconf}

\appendix

\end{document}